\documentclass[default,iicol]{sn-jnl}
\usepackage[english]{babel}
\usepackage{amssymb,amsfonts,times,amsmath,epsfig,hyperref,url,float}
\usepackage{graphicx}
\usepackage{textcomp}
\usepackage{array,multirow}
\usepackage{fmtcount}
\usepackage{booktabs}
\usepackage{pdflscape}
\usepackage{afterpage}
\usepackage{capt-of}
\usepackage{lscape}
\usepackage{longtable}
\DeclareMathOperator*{\argmin}{arg\,min}
\newcommand{\removelatexerror}{\let\@latex@error\@gobble}
\usepackage{subcaption}
\newcommand{\zf}[1]{}
\usepackage[ruled,vlined]{algorithm2e}
\usepackage{algorithmic}
\usepackage{tabularx}
\begin{document}

\title[Article Title]{Towards automated optimisation of residual convolutional neural networks for electrocardiogram classification}

\author*[1]{\fnm{Zeineb} \sur{Fki}}\email{zeinebfki@ieee.org}

\author[1]{\fnm{Boudour} \sur{Ammar}}\email{boudour.ammar@ieee.org}

\author[1,2]{\fnm{Mounir} \sur{Ben Ayed}}\email{mounir.benayed@ieee.org}

\affil*[1]{\orgname{REGIM-Lab.:REsearch Groups in Intelligent Machines, University of Sfax, National Engineering School of Sfax (ENIS)}, \orgaddress{\street{BP 1173}, \city{Sfax}, \postcode{3038}, \country{Tunisia}}}
\affil[2]{\orgname{Faculty of Science of Sfax (FSS)}, \orgaddress{\street{Road of Soukra km 4}, \city{Sfax}, \postcode{3038}, \country{Tunisia}}}


 \abstract{\textbf{Background and Objective:} The interpretation of the electrocardiogram (ECG) gives clinical information and helps in assessing heart function. There are distinct ECG patterns associated with a specific class of arrythmia. The convolutional neural network is currently one of the most commonly employed deep learning algorithms for ECG processing.  However, deep learning models require many hyperparameters to tune. Selecting an optimal or best hyperparameter for the convolutional neural network algorithm is a highly challenging task. Often, we end up tuning the model manually with different possible ranges of values until a best fit model is obtained. Automatic hyperparameters tuning using Bayesian optimisation (BO) and evolutionary algorithms can provide an effective solution to current labour-intensive manual configuration approaches.

 \textbf{Methods:} In this paper, we propose to optimise the Residual one Dimensional Convolutional Neural Network model (R-1D-CNN) at two levels. At the first level, a residual convolutional layer and one-dimensional convolutional neural layers are trained to learn patient-specific ECG features over which multilayer perceptron layers can learn to produce the final class vectors of each input. This level is manual and aims to lower the search space. The second level is automatic and based on our proposed BO-based algorithm.

 \textbf{Results:} Our proposed optimised R-1D-CNN architecture is evaluated on two publicly available ECG Datasets. Comparative experimental results demonstrate that our BO-based algorithm achieves an optimal rate of 99.95\%, while the baseline model achieves 99.70\% for the MIT-BIH database. Moreover, experiments demonstrate that the proposed architecture fine-tuned with BO achieves a higher accuracy than the other proposed architectures.

 \textbf{Conclusion:} Our optimised architecture achieves excellent results compared to previous works on benchmark datasets.}


\maketitle

\section{Introduction}\label{sec1}
The ECG is non-invasive electrical recording of the heart. The signal provides a useful information about heart health and can tell more about individuals such as gender, age, biometry and emotion recognition.
Researchers have explored this peripheral physiological signal to extract useful markers for future outcome research. Several research works have been achieved in ECG analysis. Challenges have been raised to provide an accurate ECG beats classification.
In recent years, Deep Learning (DL) or Deep Neural Network (DNN) is becoming increasingly an important research area. Typically, a DL is a neural network that contains an input layer, successive intermediate layers, namely hidden layers and an output layer. The structure of DL tries to emulate the structure of human brain given a bulky dataset, fast enough processors and a sophisticated algorithm by constructing layers of artificial neurons that can receive and transmit information.
DL offers accurate results with more training data. It is useful for unstructured data. Complex problems can be solved with a greater number of hidden layers. This structure makes possible to continuously adjust and make inferences.
It outperforms the traditional machine learning in several applications such as electrocardiogram (ECG)\cite{CAI2020103378} and Electroencephalography (EEG) \cite{rahma2020} classification, and more recently industry 4.0 \cite{ZAMORAHERNANDEZ2021103485} and COVID-19 detection \cite{ALZUBAIDI2021100025} \cite{9493719}. Recent researches provide many successful algorithms in deep learning. The Convolutional Neural Network (with the acronyms CNN) is currently one of the commonly employed deep learning algorithms for image recognition including detection of anomalies on ECG. Ebrahimi et al. \cite{EBRAHIMI2020100033} revealed that the CNN is dominantly found as the appropriate technique for feature extraction, observed in 52\% of the studies about explainable deep learning methods for ECG arrhythmia classification.
The idea of CNN comes from the biological visual cortex. The cortex consists of small regions that are sensitive to the specific areas of the visual field. Similarly, the CNN builds on small regions inside of an object that perform specific tasks. The algorithm is a hierarchical neural network. It gets the input and processes it through a series of hidden layers.  The One-Dimension Convolutional Neural Network (1D-CNN) is a distinguished variant of CNN. It is typically used for time series input with one direction x that represents the time axis.
While they have achieved excellent results in working with a variety of hard problems\cite{LI2021102203}\cite{MOITRA2020113564}, the CNNs are usually exposed to overfitting or underfitting problems. Hence, the model fails to predict the output of unseen data or even the output of the training data.
In fact, the noise introduced to the input signal slows the learning process. Various types of artifacts could lead to noisy ECG signals such as baseline wander, drift, powerline interference and muscle artifacts. The noisy signals lead to produce high false alarm rates thus the misclassification of ECG beats and misdiagnosis of cardiac arrhythmias.
In addition, the ECG signals are non-stationary. Furthermore, the high level of hyperparameters especially in fully connected layers makes the network prone to overfitting.
Several previous works have proposed methods to boost classification results based on CNN hyperparameters and regularization.
Regularizing the network structure or designing specific training schemes for stable and robust prediction is considered among the hottest topic  for efficient and robust pattern recognition in deep learning \cite{BAI2021108102}. A complex model may achieve a high performance on training data since all the inherent relations in seen data are memorized. However, the model is usually unable to perform well for unseen data including validation and test data. In order to solve this issue, different regularization methods were applied in the literature. Xu et al. \cite{XU201969} proposed SparseConnect to alleviate overfitting by sparsifying regularization on dense layers of CNNs. However, they raise, furthermore, the complexity of the model which in turn put the model harder to optimise.
In our work, we choose to randomly dropping a few nodes.
Unlike conventional methods of tuning based on manual tries to choose the best hyperparameter value, our work proposes to use BO to select an optimal configuration of dropout rate and the number of convolutional layers. In our proposal, two-level process has been established for building a robust Residual 1D-CNN (R-1D-CNN). The level one has the potential of reducing the search space of hyperparameters. The second level allows to test some configurations of the model.
The innovative contributions associated with this work can be described as follows:
\begin{enumerate}
  \item We build a novel R-1D-CNN architecture to detect features of ECG automatically. The proposed architecture presents good performance.
  \item To solve the overfitting issue and give robust classification results in real-time through automatic hyperparamters tuning, we develop an algorithm based on BO.
  \item We further explore two datasets for experimental study. Our proposal outperforms another technique of optimisation and all the previous works in ECG classification, which displays the performance of our proposed architecture.
\end{enumerate}
The rest of this paper is organized as follows:
In Section \ref{sec:background}, we outline a short background of the CNN and BO technique.
In section \ref{sec:proposed architecture}, the proposed architecture is detailed. The demonstration and performance of the proposed architecture are indicated in Section \ref{sec:experiments}. At last, Section \ref{sec:Conclusion}
concludes the paper.
\section{Background}
\label{sec:background}
\subsection{Convolutional Neural Networks}
Much of the current research on deep learning has focused on improving and validating existing deep learning algorithms rather developing new algorithms. The CNN is one of the commonly employed deep learning algorithm. A CNN learns different level of abstraction about an input. CNN performs well in image processing, including image recognition and image classification thanks to its hierarchical layers. The hierarchical property allows to increasingly learning a complex model.For instance, the model learns in the first time the basic elements, then it learns later their parts. Another advantage of CNN is the automatic extraction of feature and it requires minimal pre-processing \cite{improvedcnnhussein}.
The input of CNN is an array of pixels in the format of H xW xD where H= Height, W=Width and D=dimension. The H xW constitute the feature map and D is the depth. A grey image of size 32x32 pixels is represented by an array 32X32X1 while an RGB image of the same size is represented by an array of 32X32X3.
The structure may include convolutional layers hence its name, pooling layers, Rectified Linear Unit (ReLU) layers and fully connected layers.
\begin{itemize}
  \item Convolutional layer:  the main layer of CNN. It consists of a set of filters that exploits the local spatial correlation assuming that near pixels are more correlated than distant pixels. The size of the filter defines the size of each feature map and its depth defines the number of feature maps. All local regions share the same weights called weight sharing.
Mathematically saying, a convolution acts as a mixer, mixing two functions to obtain a reduced data space while preserving the information. The model involves training a multilayer architecture without the explicit need of handcrafted input features and is able to extract automatically the features such as edge, blur and sharpen. It helps to remove noise.
  \item Pooling layer:  common use is the max-pooling, which implements a sliding window. The max-pooling operation slides over the layer and takes the maximum of each region with a step of stride vertically and horizontally.
  \item ReLU: is a non-linear activation function. It performs a threshold operation. The output takes the same value as the input for the positive values and zero otherwise. The function is used by default for many deep learning algorithms since it performs well and avoids vanishing problem.
  \item Fully connected or dense layer: in a fully connected layer, every neuron is connected to every neuron in the next layer. A model may contain one or more fully connected layer. The dense layer can be the last layer for the classification.
\end{itemize}
Based on the output, different convolutional dimensions can be used. 1D-CNN is typically used for time series input with one direction x that represents the time axis. Common uses of 1D-CNN are proposed for ECG data classification and anomaly detection \cite{CHEN2021}. The output shape is one dimension. 2D-CNN  performs well for image recognition and classification as the input is an image is of 2 dimensions. The output shape is 2 dimensions. The convolution is calculated based on two directions (x,y). With increasing number of dimensions, CNN 3D applies a three dimensions filter. The filter moves in three directions (x, y, z). The model is helpful in drug discovery \cite{9305294}.\\
\subsection{Bayesian optimisation}
\label{sec:hyperparameter}
The effective use of machine learning algorithms is associated with hyperparameters tuning. The hyperparameters adjust the model to a specific database and avoid ongoing training costs.
To get up speed on hyperparameters tuning, BO can be used. The technique is based on Bayes' theorem \cite{BERRAR2019403} to select the best configuration of hyperparameter values. The Bayes' theorem consists of calculating the conditional probability of an event.
Bayes'theorem uses prior probability distributions to be able to produce posterior probabilities. Prior probability could be the probability of an event before new knowledge is collected.
The probability of A conditional on B is defined as \ref{eq:bayes theorem}.
\begin{equation} \label{eq:bayes theorem}
P(A \mid B) = \frac{P(A \bigcap B)}{P(B)}= \frac{P(A)*P(B \mid A)}{P(B)}
\end{equation}
Where:\\
  $P(A)=$ The probability of A occurring\\
  $P(B)=$ The probability of B occurring\\
  $P(A \mid B)=$ The probability of A given B\\
  $P(B \mid A)=$ The probability of B given A\\
  $P(A \bigcap B)=$ The probability of both A and B occurring\\
BO provides a global optimal solution. By limiting the hyperparameters search space on ranges of values, the algorithm develops a probabilistic model of the objective function named the surrogate function.
Mathematically saying, the algorithm is interested in solving equation \ref{eq}:\\
\begin{equation} \label{eq}
x^{*} =\argmin_x f(x)
\end{equation}
This optimisation method takes into account the problem of noise present in the evaluations of equation \ref{eq1}
\begin{equation} \label{eq1}
y=f(x)
\end{equation}
Where $f$ is a black box and expensive to evaluate.\\
Starting from default parameters e.g. parameter ranges that are used in the literature, the performance evaluation calculated using a numeric score or cost such as the accuracy rate. The aim is selecting a best configuration that maximize or minimize the cost. The best result achieved by a couple of hyperparameters would be used to construct the tuned model. Hence, the hyperparameters are assigned.
For more details about hyperparameter optimisation for machine learning models based on BO, please see \cite{Bayesian-2020}.
\section{Methods}
\label{sec:proposed architecture}
The proposed classifier is based on CNN algorithm.  The first setting of hyperparameters is done manually. The process is iterative to accomplish an acceptable rate of accuracy. We add layers and nodes to the model gradually. The increasing layer number made the manual optimisation harder. This configuration is given to the optimisation algorithms as the default parameters and runs as the first iteration. By optimising the neural network loss, the smoothing parameters are optimised to perform the prediction task.
A novel R-1D-CNN architecture is presented in our work. The optimisation method is described below.
\begin{figure*}[!t]
\centering
\includegraphics[width=0.8\textwidth]{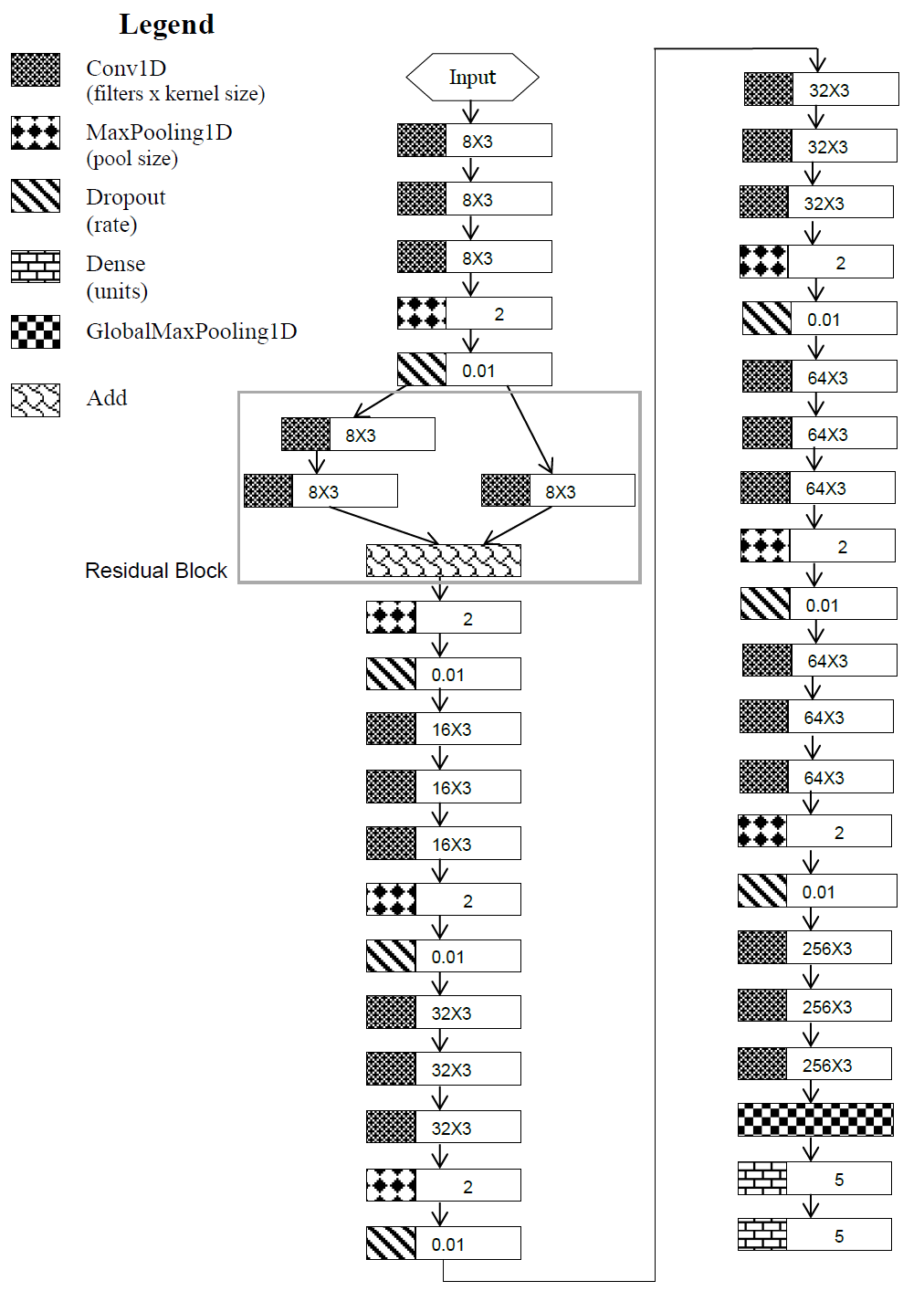}
\caption{Overall structure of our proposed architecture- level 1}
\label{fig:model}
\end{figure*}
\subsection{First Level: Architecture Building}
\label{first level}
Our proposed CNN at level one  was created with 41 plain hidden layers, the first five layers consisting of three convolutional layers, one maxpooling layer and a dropout layer. The block of five layers is repeated seven times with different filter size. The final block of layers consists of three convolutional layers connected to max pooling layer which is followed by another three dense layers.
As the network becomes more and more deeper, residual connection \cite{residual2016} is introduced. Hence a new level of depth is appeared. Deep residual learning comes with the benefit of solving the issue involved with vanishing/exploding gradients as well as degradation problem.
Residual network achieves this by employing skip connections, or shortcuts to leap across a number of layers.
The number of residual blocks in our architecture is fixed to one. The skip connection is located in the position displayed as a residual block in Figure \ref{fig:model}. This new architecture allows an efficient training by including skip connections.
\subsection{Second Level: Architecture optimisation}
To enhance the architecture performance and avoid overfitting issue, we choose to use the Bayesian optimiser.
The last is made using Bayesian inference and Gaussian process (GP). This approach is an appropriate algorithm to optimise hyperparameters of classification. By choosing which variables to optimise, and specifying the ranges to search in, the algorithm selects the optimal values.
The GP is a well known surrogate model for BO employed for approximating the objective function. It performs well in small dimensional spaces specifically when the number of features meets five features.  Table \ref{tab:searchSpace} illustrates the selecting hyperparameters, their type and their ranges.
 A deep learning model is constructed
according to the first level, and the most likely point to be maximized by acquisition function is identified. Some hyperparameters that are very responsive to changes are chosen at first level such as learning rate.
BO enables fine-tuning of the model through the regularization of the penalty and determining the optimal number of layers. We choose dropout layer as a technique of regularization.
\begin{table*}[t]
  \centering
\caption{Hyperparameters end up being run optimisation}
\label{tab:searchSpace}
\begin{tabular}{@{}llll@{}}
\toprule
Hyperparameter &  Type &  \multicolumn{2}{c}{Ranges}\\
 &   &  Low & High \\
\midrule
Drop Rate & Real & 1e-2 & 1e-1\\
Number of Dense Layers & Integer & 1 & 6\\
Number of Convolutional Layers & Integer & 1 & 6\\
Learning Rate & Real & 1e-3 & 1e-1\\
Adam Decay & Real & 1e-6 & 1e-5\\
\botrule
\end{tabular}
\end{table*}
\section{Results}
\label{sec:experiments}

\subsection{Setup}
We used the python and its data science library to implement our algorithms. We implemented the algorithms using the Keras of TensorFlow library version 2.5 on a Tesla P100 GPU and 25 GB that are provided by  Google Colaboratory Notebooks.
The training set contains 70\% of randomly selected beats and the rest is divided into test and validation set. Each set contains 50 \% of the remained beats.
\subsection{Database}
Our proposed model is trained on two publicly available datasets: (1) The MIT-BIH dataset \cite{mitbih} that includes 48 ECG recordings of 30 minutes duration of 47 subjects and 250 sampling rate. Each record is annotated by specialists and can be utilized as ground truth for training, validation and test. The collected data is preprocessed. We build a new dataset that consists of 82813 segments.
For beat segmentation, we consider a fixed window multiple of frequency. The raw ECG signal doesn't require any pre-filtering technique or feature extraction step as used in traditional machine learning algorithms. The database is relatively noise free. Furthermore, the CNN is robust to the noise and features are automatically extracted during the learning process. (2) The second dataset is 10,000 ECG patients dataset \cite{database-10000-patient}. This dataset consists of 10646 subjects of 10 seconds duration and 500 Hz sampling rate.
\subsection{Performance}
\begin{figure}[h]
\centering
\includegraphics[width=0.5\textwidth]{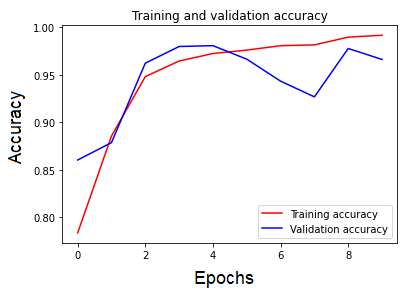}
\caption{Training and Validation Accuracy at level one}
\label{fig:accuracy}
\end{figure}

\begin{figure}[h]
\centering
\includegraphics[width=0.5\textwidth]{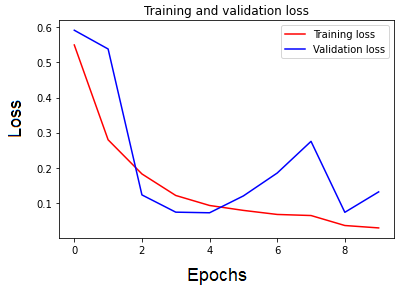}
\caption{Training and Validation Loss at level one}
\label{fig:loss}
\end{figure}

The performance of our proposed model was evaluated using seven experiments. The first experiment is used to build an architecture to fit MIT-BIH dataset and is achieved 99.70\%.
The figures \ref{fig:accuracy} and \ref{fig:loss} illustrate the accuracy and loss obtained during the training phase. The model runs 100 epochs with early stopped enabled. While the size of the input vector and the number of the hidden layers is large, the model converges in an extremely small time (8 epochs).\\
However, the gap between the validation and the training is significant.\\
At level two, we introduce the BO. A form of pseudo-code is written to provide precise descriptions of what BO does. The pseudo-code is presented in Algorithm \ref{alg:bayesian}.
The performance of the algorithm has increased. The BO produced an improvement right after the \ordinalnum{13} iteration. The numerical experiments are showing that the resulting accuracy for the optimisation with a finite budget outperforms the accuracy of the baseline model.

\SetKwInput{kwInit}{Initialization}
\SetKw{KwBy}{by}

\begin{figure}[h]
\centering
\includegraphics[width=0.5\textwidth]{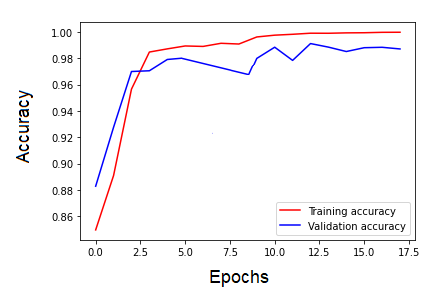}
\caption{Training and Validation Accuracy - Optimised model}
\label{fig:accuracyFinal}
\end{figure}

\begin{figure}[h]
\centering
\includegraphics[width=0.5\textwidth]{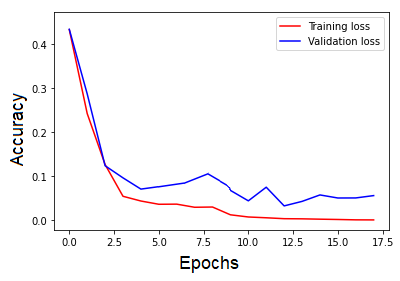}
\caption{Training and Validation Loss - Optimised model}
\label{fig:lossFinal}
\end{figure}

\begin{figure}[h]
\centering
\includegraphics[width=0.5\textwidth]{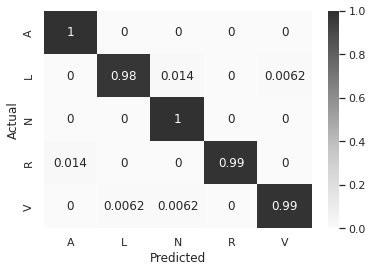}
\caption{
Normalized Confusion Matrix for identification of five classes}
\label{fig:confusionmatrix}
\end{figure}

\begin{algorithm*}

\LinesNumbered
\SetKwBlock{Begin}{Begin}{}
\SetAlgoLined
  \Begin{
  \SetAlgoLined
\KwIn{\ Training set $D=\left\{(\mathbf{x}_1,y_1),(\mathbf{x}_2,y_2),\dots,(\mathbf{x}_m,y_m)\right\}$\;
				Default hyperparameters set $H=\left\{(h_1,h_2,\dots,h_d)\right\}$\; }
\KwOut{\ The optimal hyperparameters producing the best classification accuracy}
\kwInit{$optimisationBudget=15$}
 \tcp*{The number of evaluations of fitness function}
 \SetKwFunction{FcreateModel}{createModel}
 \SetKwFunction{Ffitness}{fitness}
 \While{$i\leq optimisationBudget$}{
  $gaussianProcessesOptimise(func=fitness,H)$\;

 }
\SetKwProg{Fn}{Function}{:}{\KwRet $model$}
  \Fn{\FcreateModel{$dropRate$, $numberDenseLayers$, $numberConvLayers$, $learningRate$, $adamDecay$}}{
        $model$=$new Model()$\;
        \For{$i\gets1$ \KwTo $numberConvLayers$ \KwBy $1$}{
          $model$.add($convLayer$)\;
         }
         $model$.$add$($maxPoolingLayer$)\;
         $model$.$add$($dropOutLayer(dropRate)$)\;
         $interimLayerIn$=$interimLayerOut$=$convLayer$\;
         \For{$i\gets1$ \KwTo $numberConvLayers$ \KwBy $1$}{
          $interimLayerOut$=$convLayer$($interimLayerOut$)\;
         }
         $residualLayer$=$add$($interimLayerIn$,$interimLayerOut$)\;
         $model$.$add$($residualLayer$)\;
         $model$.$add$($maxPoolingLayer$)\;
         $model$.$add$($dropOutLayer(dropRate)$)\;
         \For{$i\gets1$ \KwTo $5$ \KwBy $1$}{
          \For{$j\gets1$ \KwTo $numberConvLayers$ \KwBy $1$}{
          $model$.add($convLayer$)\;
         }
         $model$.$add$($maxPoolingLayer$)\;
         $model$.$add$($dropOutLayer(dropRate)$)\;
         }
         \For{$i\gets1$ \KwTo $numberDenseLayers$ \KwBy $1$}{
          $model$.add($denseLayer$)\;
         }
         $model$.$add$($denseLayer(5,activation=softmax)$)\;
         $opt$ = $optimisers.Adam(learning_rate=learningRate, decay= adamDecay)$\;
        $model$.$compile(optimiser=opt, loss=losses.categoricalCrossEntropy, metrics=['acc'])$\;
         \SetKwProg{Fn}{Function}{:}{}
  \Fn{\Ffitness{$dropRate$, $numberDenseLayers$, $numberConvLayers$, $learningRate$, $adamDecay$}}{
        $model$=\FcreateModel{$dropRate$, $numberDenseLayers$, $numberConvLayers$, $learningRate$, $adamDecay$}\;
        $result=model.fit$\;
  }
  }
 \caption{Algorithm BO-based pseudo-code}
 \label{alg:bayesian}
  }
\end{algorithm*}
Finally, we build the optimised model for test. The figures \ref{fig:accuracyFinal} and \ref{fig:lossFinal} illustrate the accuracy and loss obtained during the training phase. The training accuracy and training loss are respectively close to the validation accuracy and validation loss at second level.
The confusion matrix is displayed in Figure \ref{fig:confusionmatrix}. Typically, the diagonal elements present the rate of items that are well predicted. Off-diagonal items are mislabelled. The proposed optimised R-1D-CNN properly predicted ECG signals of
five distinct classes with a high accuracy of 99.95\%. By reviewing the individual performance ratios of the different classes, we notice that the minimum recognition rate belongs to Left bundle branch block class (L) with 98\%. The highest recognition performance is in the Atrial premature (A) and Normal (N) classes with 100\%. The Right bundle branch block (R) and the Premature ventricular contraction (V) achieve 99\%.
The experiments 4-7 aims at testing the model on different benchmark datasets and using two different optimisation techniques. Other metrics are used to evaluate the classification performance. The Recall, the Precision and F1 score can be calculated respectively by Equations (\ref{eq:recall})–(\ref{eq:f1})
\begin{equation}\label{eq:recall}
  Sensitivity (\%) = Recall = \frac{TP}{TP+FN}\times 100
\end{equation}
\begin{equation}\label{eq:precision}
  Precision (\%) = \frac{TP}{TP+FP}\times 100
\end{equation}

\begin{equation}\label{eq:f1}
\begin{split}
F1 (\%) = \frac{2*Precision*Recall}{Precision+Recall} \\
= \frac{2*TP}{2*TP+FP+FN}\times 100
\end{split}
\end{equation}
Table \ref{tab:performance-different-datasets} display the  achieved results. According to the table, the BO outperforms the PSO. The model fit well for the MIT-BIH and 10,000 patients databases.\\
\begin{table*}[!t]
    \centering
    \caption{Performance classification using different optimisation techniques}
    \label{tab:performance-different-datasets}
    \begin{tabular}{lccccc}
        \toprule
         Dataset&Performance &Without optimisation&With Bayesian&With PSO\\
        \midrule

        \multirow{3}{*}{MIT-BIH}     &  \multirow{3}{*}{Precision}       &  \multirow{3}{*}{99.7} & \multirow{3}{*}{99.95}       &  \multirow{3}{*}{99.8}                \\
        &&&&\\
        \cline{2-5}
                 & Recall        & 99.02       &99.23       & 99.06                \\
        &&&&\\
         \cline{2-5}
              & F1 score        & 99.35       &99.58       & 99.42              \\
         &&&&\\
          \hline
         \multirow{3}{*}{10,000 Patients}     &  \multirow{3}{*}{Accuracy}       &  \multirow{3}{*}{98.33} & \multirow{3}{*}{99.92}       &  \multirow{3}{*}{99.27}                \\
        &&&&\\
        \cline{2-5}
                 & Recall        & 98.04        &99.22       & 99.15               \\
        &&&&\\
         \cline{2-5}
              & F1 score        & 98.18        &99.4       & 99.2              \\
         &&&&\\

        \bottomrule
    \end{tabular}
\end{table*}

\subsection{Discussion}
\begin{table*}[t]
  \centering
\caption{Classification accuracy of different architectures}
\label{tab:comparison}
\begin{tabular}{*{9}{l}}
\toprule
Reference&Year&Database& Architecture + Classifier&Accuracy (\%)\\
\midrule
\cite{Yildirim-LSTM-CMPB-2019}&2019&MIT-BIH&Autoencoder + LSTM&99\\
\cite{NURMAINI2021100507}&2020&MIT-BIH&Bidirectional LSTM&99.64\\
\cite{RATH2021102820}&2021&MIT-BIH&GAN+LSTM&99.2\\
\cite{MURAT2021107473}&2021&10,000 Patients&CNN&98\\
\cite{2022-sciencedirect-CNN-ComputerMethods}&2022&MIT-BIH&Improved deep&88.99\\
&&&residual CNN&\\
\textbf{The proposed}&\textbf{2022}&\textbf{MIT-BIH}&\textbf{R-1D-CNN}&$>\textbf{99.9}$\\
\textbf{Bayesian-optimised}&&\textbf{and}&&\\
\textbf{R-1D-CNN}&&\textbf{10,000 Patients}&&\\
\botrule
\end{tabular}
\end{table*}

\begin{table*}[!h]
\footnotesize
\begin{center}
\begin{minipage}{\textwidth}
\caption{One-dimensional CNN optimised by BO: performance comparison of the proposed method with other methods}
\label{tab:bayesian}
\begin{tabular*}{\textwidth}{@{\extracolsep{\fill}}|c|c|c|c|c|c|c|c|c|c|@{\extracolsep{\fill}}}
\toprule
Method&Year&DataBase&Search Space&Ranges&Optimal&$n$&Epochs&\multicolumn{2}{c|}{Accuracy (\%)} \\
 & & & & &Value&&&Baseline&Optimal\\
\midrule
\multirow{10}{*}{\cite{SAMEEN2020104249}}&\multirow{10}{*}{2020}&\multirow{10}{*}{Landslide}&Number of filters&4-512&117&\multirow{10}{*}{25}&\multirow{10}{*}{63}&\multirow{10}{*}{80.50}& \multirow{10}{*}{83.50}\\
&&&Sequence length&[3, 5, 10, 12]&10&&&&\\
&&&Batch size&4-128&16&&&&\\
&&&Activation function&[ReLU, Linear,&ELU&&&&\\
&&&&Tanh, ELU,&&&&&\\
&&&&Sigmoid]&&&&&\\
&&data&optimisation method&[SGD, Adam,&Adagrad&&&&\\
&&&&Adamax,Adadelta,&&&&&\\
&&&&Adagrad, RMSprop,&&&&&\\
&&&&Nadam]&&&&&\\
&&&Neurons in&4-512&240&&&&\\
&&&hidden layers&&&&&&\\
&&&Dropout rate&0-0.8&0.66&&&&\\
\hline
\multirow{14}{*}{\cite{transp_cnn1d_baye_2020}}&\multirow{14}{*}{2020}&\multirow{14}{*}{Acceleration}&Learning rate
&$e^{-5}-e^{-2}$&$e^{-5}$&\multirow{14}{*}{150}&\multirow{14}{*}{54}&\multirow{14}{*}{84.75}&\multirow{14}{*}{93.53}\\
&&&Batch size&16–64&17&&&&\\
&&&Epoch&20–60&54&&&&\\
&&&Layer number&1–5&4&&&&\\
&&&Conv1 filter number&6–16&10&&&&\\
&&&Conv1 filter size&3–11&4&&&&\\
&&&Conv2 filter number&12–32&27&&&&\\
&&&Conv2 filter size&3–11&11&&&&\\
&&signals&Conv3 filter number&24–64&24&&&&\\
&&&Conv3 filter size&3–11&10&&&&\\
&&&Conv4 filter number&48–128&66&&&&\\
&&&Conv4 filter size&3–11&3&&&&\\
&&&Conv5 filter number&96–256&-&&&&\\
&&&Conv5 filter size&3–11&-&&&&\\
\hline
\multirow{9}{*}{\cite{app11104660}}&\multirow{9}{*}{2021}&\multirow{9}{*}{UrbanSound8K}&Number of filters&16-512&158 &\multirow{9}{*}{25}&\multirow{9}{*}{50}& \multirow{9}{*}{92.4}&\multirow{9}{*}{94.4}\\
&&&Kernel size&[2, 4, 6-12]&3&&&&\\
&&&Batch size&16-256&64&&&&\\
&&&Activation function&[ReLU, Linear,&&&&&\\
&&&&Sigmoid, Tanh]&-&&&&\\
&&dataset&optimisation method&[Adam, EAG,&Adam&&&&\\
&&&&RMSprop, Nadam,&&&&&\\
&&&&Adamax, Adadelta,&&&&&\\
&&&& Adagrad]&&&&&\\
&&&Dropout rate&0-0.5&0.266&&&&\\
\hline
\multirow{12}{*}{\textbf{Proposal}}&\multirow{12}{*}{\textbf{2022}}&\multirow{12}{*}{\textbf{MIT-BIH/}}&\textbf{Drop rate}&$e^{-2}-e^{-1}$&\textbf{0.010738/ 	}&\multirow{12}{*}{\textbf{15}}&\multirow{12}{*}{\textbf{8}}& \multirow{12}{*}{\textbf{99.70}}&\multirow{12}{*}{\textbf{$>$99.9}}\\
&&&&&\textbf{ 0.011315}&&&&\\
&&&\textbf{Number of}&&&&&&\\
&&&\textbf{dense layers}&\textbf{1-6}&\textbf{1/}&&&&\\
&&&&&\textbf{6}&&&&\\
&&&\textbf{Number of}&&&&&&\\
&&&\textbf{convolutional layers}&\textbf{1-6}&\textbf{3/}&&&&\\
&&&&&\textbf{2}&&&&\\
&&\textbf{10,000Patients}&\textbf{Learning rate}&\textbf{$e^{-3}-e^{-1}$}&\textbf{0.001832/}&&&&\\
&&&&&\textbf{0.0017}&&&&\\
&&&\textbf{Adam decay}&$e^{-6}-e^{-5}$&\textbf{0.000006/}&&&&\\
&&&&&\textbf{0.000004}&&&&\\
\botrule
\end{tabular*}
\end{minipage}
\end{center}
\end{table*}
Both of our architectures at level one and level two present novelties. We demonstrate that the proposed R-1D-CNN architecture, fine tuned by BO method is efficient compared to the state-of-the art architectures. In our experiments,  we exploit the R-1D-CNN to classify the ECG signal of two databases. Table \ref{tab:comparison} displays the classification accuracy of different architectures. A prior study has shown that ECG signals have successfully learned by AutoEncoder Long Short-Term Memory (AE-LSTM). Other recently developed convolutional architectures have been trained on similar databases. Nevertheless, our proposal achieves the highest accuracy.\\
Compared to the previous works on MIT-BIH database, the CNN architecture provides an automatic extraction of the features, and does not require pre selection feature step. Hence, it is more generic. This attractive feature may facilitate the application of the model in more real-time context,  such as the Internet of Things (IoT) data analysis.
The results obtained by our optimised architecture are in agreement with those obtained with \cite{transp_cnn1d_baye_2020} \cite{SAMEEN2020104249} \cite{app11104660}.
Table \ref{tab:bayesian} presents different approaches worked on 1D-CNN and optimised by BO. The table presents mainly the selected hyperparameters, their search space, the obtained optimal values that are used in BO.
Our model is able to reach a high classification rate for different benchmark datasets with a minimum optimisation budget and variables. For instance, the drop rate with value 0.010738 and 0.011315 is the optimised value to fit the MIT-BIH database and 10,000 Patients database, respectively.

\section{Conclusion}
\label{sec:Conclusion}
In this paper we address optimisation challenges for the R-1D-CNN model and propose a novel architecture for ECG analysis. In addition, we develop an algorithm based on BO to produce robust classification results in real time through automatic hyperparamters tuning.
Comparative experimental results performed on two publicly available ECG Datasets demonstrate that the our BO-based algorithm can outperform
the state-of-art approaches. The BO achieves for instance an optimal rate of 99.95\%, while the baseline model achieves 99.70\% for the MIT-BIH database.
In future, we plan to test the algorithms on other databases, especially for dialysis applications. We will also will introduce others type of layers and classifiers to manage and optimise the complexity of the network.

\backmatter



\section*{Compliance with Ethical Standards}

\begin{itemize}
\item Funding: The research leading to these results has received funding from the Ministry of Higher Education and Scientific Research of Tunisia under the grant agreement number LR11ES48.
\item Disclosure of potential conflicts of interest:
The authors declare no conflict of interest.
\item Ethics approval: This article does not contain any studies with human participants or animals performed by any of the authors.
\item Informed consent: Not applicable
\item Consent to participate: Not applicable
\item Consent for publication: Not applicable
\item Availability of data and materials:
The ECG signals are obtained from the MIT-BIH arrhythmia database and 12-lead electrocardiogram database for arrhythmia research covering more than 10,000 patients. All the databases are public  and available online:\\
Link to MIT-BIH  arrhythmia database:\\
\url{https://physionet.org/content/mitdb/1.0.0/} \\
The original publication is referenced by \cite{mitbih}.\\
Link to the 12-lead electrocardiogram database:\\
\url{https://doi.org/10.6084/m9.figshare.c.4560497.v2}
The original publication is referenced by \cite{database-10000-patient}.
\item Authors' contributions:
Zeineb Fki: developed the method, collected the data,
performed the experiments and drafted the manuscript.\\
Boudour Ammar: revised the software, interpreted the
results and supervised the project.\\ Mounir Ben Ayed:conceived the study and supervised the project. All authors
read and approved the final manuscript.
\end{itemize}

\noindent
\bibliographystyle{unsrtnat}
\bibliography{BibZeineb}
\end{document}